\documentclass[final,3p,times,twocolumn]{elsarticle}

\usepackage{amssymb}

\usepackage{amsmath}

\usepackage{color}
\usepackage{hyperref}
\hypersetup{colorlinks, linkcolor=blue}

\makeatletter
\def\ps@pprintTitle{%
  \let\@oddhead\@empty
  \let\@evenhead\@empty
  \let\@oddfoot\@empty
  \let\@evenfoot\@oddfoot
}
\makeatother

\begin{document}

\begin{frontmatter}



\title{Soliton compression and supercontinuum spectra in nonlinear diamond photonics}


\author[inst1,inst2]{O. Melchert}
\author[inst2,inst4]{S. Kinnewig}
\author[inst3]{F. Dencker}
\author[inst1]{D. Perevoznik}
\author[inst1,inst2]{S. Willms}
\author[inst1,inst2]{I. Babushkin}
\author[inst3]{M. Wurz}
\author[inst2,inst5]{M. Kues}
\author[inst2,inst4]{S. Beuchler}
\author[inst2,inst4]{T. Wick}
\author[inst1,inst2]{U. Morgner}
\author[inst1,inst2]{A. Demircan}

\affiliation[inst1]{organization={Leibniz Universität Hannover, Institute of Quantum Optics}, 
            addressline={Welfengarten 1}, 
            city={Hannover},
            postcode={30167}, 
            country={Germany}}

\affiliation[inst2]{organization={Leibniz Universität Hannover, Cluster of Excellence PhoenixD}, 
            addressline={Welfengarten 1A}, 
            city={Hannover},
            postcode={30167}, 
            country={Germany}}

\affiliation[inst4]{organization={Leibniz Universität Hannover, Institute of Applied Mathematics}, 
            addressline={Welfengarten 1}, 
            city={Hannover},
            postcode={30167}, 
            country={Germany}}

\affiliation[inst3]{organization={Leibniz Universität Hannover, Institute of Micro Production Technology}, 
            addressline={An der Universität 2}, 
            city={Garbsen},
            postcode={30823}, 
            country={Germany}}

\affiliation[inst5]{organization={Leibniz Universität Hannover, Institute of Photonics}, 
            addressline={Nienburger Str.~17}, 
            city={Hannover},
            postcode={30167}, 
            country={Germany}}

\begin{abstract}
We numerically explore synthetic crystal diamond for realizing novel light
sources in ranges which are up to now difficult to achieve with other
materials, such as sub-10-fs pulse durations and challenging spectral ranges.  
We assess the performance of on-chip diamond waveguides for
controlling light generation by means of nonlinear soliton dynamics.  
Tailoring the cross-section of such diamond waveguides allows to design
dispersion profiles with custom zero-dispersion points and anomalous dispersion
ranges exceeding an octave.
%
%
%
Various propagation dynamics, including supercontinuum generation by soliton
fission, can be realized in diamond photonics.
In stark contrast to usual silica-based optical fibers, where such processes
occur on the scale of meters, in diamond millimeter-scale propagation distances are
sufficient.
Unperturbed soliton-dynamics prior to soliton fission allow to identify a pulse
self-compression scenario that promises record-breaking compression factors on
chip-size propagation lengths.
\end{abstract}





\end{frontmatter}


\section{Introduction}
\label{sec:intro}

Nonlinear diamond photonics provides an attractive technical basis for on-chip
photonic applications \cite{Hausmann:N:2014}, and has triggered numerous
research efforts in recent years. 
Owing to the unique material properties of diamond
\cite{Rath:PSS:2015,Gaeta:NP:2019}, given by its large Kerr nonlinearity, wide
bandgap, high refractive index, negligible multi-photon loss, and transmission
window spanning from the ultrarviolet to the far-infrared, impressive
demonstrations of photonic devices with novel functionalities have emerged. 
This includes, e.g., its use as a platform for quantum communication
\cite{Beveratos:PRL:2002}, and integrated high-$Q$ optical resonators
\cite{Hausmann:NL:2013,Hausmann:N:2014}, operating at new wavelengths compared
to existing chip-based nonlinear photonic devices for frequency comb generation
\cite{Kippenberg:S:2011,Gaeta:NP:2019}.
It thus exceeds its use in quantum optics applications and is becoming a
versatile material for optical devices.
A direct transfer of concepts from photonic crystal fibers and
silicon-based waveguides \cite{Ding:OE:2008,BlancoRedondo:NC:2014}, such as,
e.g., pulse-compression schemes and soliton-effects, to the diamond-based
platform seems possible.

Here, we consider the supercontinuum generation process
\cite{Ranka:OL:2000,Agrawal:BOOK:2019,Mitschke:BOOK:2016,Skryabin:RMP:2010}, a
paradigm of optical pulse propagation in fibers, which has revolutionized
optical coherence tomography \cite{Hartl:OL:2001}, and frequency metrology
\cite{Udem:N:2002}. In common silica-based photonic crystal fibers, this
process occurs on the lengthscale of several centimeters
\cite{Dudley:RMP:2009}, or even meters \cite{Ranka:OL:2000}.
We use the propagation properties of a diamond waveguide surrounded by silica
\cite{Hausmann:N:2014}, and demonstrate in terms of numerical simulations that
the supercontinuum generation process unfolds on a much shorter,
millimeter-length propagation scale.
In our analysis, 
we investigate the
propagation dynamics of ultrashort optical pulses via the generalized
nonlinear Schrödinger equation \cite{Agrawal:BOOK:2019}, taking into account
higher-order dispersion, pulse self-steepening
\cite{deMartini:PR:1967,deOliveira:JOSAB:1992}, and the Raman effect
\cite{Gordon:OL:1986}. 
This accounts for various processes that support the generation of widely
extended supercontinuum spectra, such as the modulation instability
\cite{Demircan:OC:2005}, soliton-fission
\cite{Husakou:PRL:2001,Demircan:APB:2007}, and self-frequency shift of Raman
solitons \cite{Gordon:OL:1986}. 
The initial stage of the supercontinuum generation process allows to identify a
pulse self-compression mechanism based solely on soliton-effects
\cite{Mollenauer:OL:1983,Oliver:OL:2021}.
Exploiting this mechanism for higher-order soliton compression, we achieve
record-breaking pulse-compression factors, outperforming recent studies in
silicon-nitride waveguides \cite{Oliver:OL:2021}.  
For instance, the compression of a hyperbolic-secant shaped input pulse of
$300\,\mathrm{fs}$, corresponding to a higher-order soliton of order $N=15$,
down to $5.4\,\mathrm{fs}$ is achieved on a propagation length of only
$6.33\,\mathrm{mm}$.  
In this respect, diamond allows to consider comparatively high pulse
intensities enabling conditions that facilitate high-order soliton propagation
effects when pumping in the domain of anomalous dispersion.
The fabrication of diamond waveguides with cross-sections that allow to engineer
the required dispersion profiles, working at telecom wavelengths and exhibiting
the key-feature of a wide domain of anomalous dispersion, is technically
feasible \cite{Hausmann:N:2014,Feigel:OL:2017}. 
In this regard, since silica-based fibers have clear limitations concerning
transparency and convenient dispersion profiles (as described in Sect.\
\ref{sec:methods} below), 
working with diamond seems beneficial,
e.g., the ability to engineer unusual dispersion profiles with several
zero-dispersion points leads to the observation of new phenomena
\cite{Melchert:PRL:2019,Tam:PRA:2020,Tsoy:PRA:2007,Melchert:OL:2021,Willms:PRA:2022}.
%

In Sect.\ \ref{sec:methods} we introduce the numerical model for 
nonlinear pulse propagation in more detail.
Section \ref{sec:results} contains the analysis of the supercontinuum
generation process and the pulse self-compression scheme in the considered
diamond waveguide.
Finally, we discuss our results and conclude in Sect.\ \ref{sec:discussion}. 

\section{Methods}
\label{sec:methods}


For the numerical simulation and analysis of the nonlinear $z$-propagation
dynamics of ultrashort laser pulses we use the generalized nonlinear
Schrödinger equation (GNLS) \cite{Agrawal:BOOK:2019,Dudley:RMP:2009}
\begin{align}
\partial_z A = i \sum_{n\geq 2}& \frac{\beta_n}{n!}(i\partial_t)^n A + i \gamma \left( 1+i\omega_0^{-1}\partial_t\right) \notag \\
&\times \left[\,A(z,t)\int R(t^\prime) |A(z,t-t^\prime)|^2~{\rm{d}}t^\prime\right],
                                   \label{eq:GNLS}
\end{align}
for a complex-valued field $A\equiv A(z,t)$ on a periodic time-domain of extend
$T$ with boundary condition $A(z,-T/2)=A(z,T/2)$. 
%
%
%
In Eq.~(\ref{eq:GNLS}), $t$ is a retarded time measured in a reference frame
moving with the group velocity at $\omega_0$, where $\omega_0$ is a reference
frequency with units $\mathrm{rad/ps}$.
The real-valued coefficients $\beta_n$ specify the dispersion coefficients of
order $n$ with units $\mathrm{ps}^n/\mathrm{m}$, and $\gamma$ specifies the
nonlinear coefficient with units $\mathrm{W^{-1}/m}$.

\begin{figure}[t!]
\includegraphics[width=\linewidth]{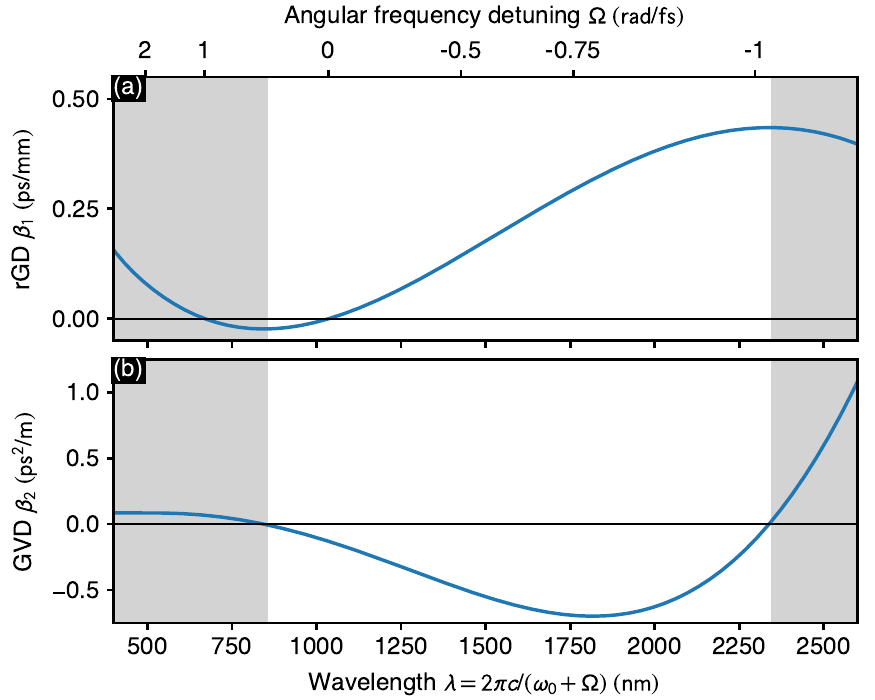}
\caption{Characteristics of the propagation constant of the considered diamond 
waveguide, reproduced following Fig.~5(a) Ref.~\cite{Hausmann:N:2014}.  (a)
Frequency dependence of the relative group delay (rGD). (b) Frequency
dependence of the group-velocity dispersion (GVD), with zero-dispersion points
are at $\lambda_{\mathrm{Z1}}\approx 843\,\mathrm{nm}$, and
$\lambda_{\rm{Z2}}\approx 2340\,\mathrm{nm}$.
Domains of normal dispersion are shaded gray. 
Top axis in (a) indicates the detuning $\Omega$, related to the wavelength
through $\lambda = 2 \pi c/(\omega_0+\Omega)$ with speed of light $c$.
\label{fig:OM:01}}
\end{figure}

To model dispersive and nonlinear effects in diamond waveguides we use a
propagation constant $\beta(\Omega)=\sum_{n\geq 2} (\beta_n/n!) \Omega^n$,
where $\Omega=\omega-\omega_0$ defines an angular frequency detuning,
characterized by the relative group delay
$\beta_1(\Omega)=\partial_\Omega\,\beta(\Omega)$ shown in
Fig.~\ref{fig:OM:01}(a), and group-velocity dispersion
$\beta_2(\Omega)=\partial_\Omega^2\,\beta(\Omega)$ shown in
Fig.~\ref{fig:OM:01}(b).
This broadband anomalous dispersion profile characterizes a silica embedded
diamond waveguide with height $H=950\,\mathrm{nm}$ and width
$W=875\,\mathrm{nm}$, extracted from Ref.~\cite{Hausmann:N:2014}.
This waveguide device was designed for the telecom wavelength range and
exhibits a wide domain of anomalous dispersion, bounded by zero dispersion
points at $\lambda_{\mathrm{Z1}}\approx 843\,\mathrm{nm}$ and
$\lambda_{\rm{Z2}}\approx 2340\,\mathrm{nm}$, see Fig.~\ref{fig:OM:01}.
We further use $\gamma=9.6\,\mathrm{W^{-1}/m}$ \cite{Feigel:OL:2017}.
The Raman effect is included via the total response function  
\begin{align}
R(t)=(1-f_{{R}})\,\delta(t) + f_{{R}}\,h_{{R}}(t), \label{eq:R}
\end{align}
where the first term defines the instantaneous Kerr response, and where the
second term specifies a generic two-parameter Raman response function
\cite{Blow:JQE:1989,Stolen:JOSAB:1989}
\begin{align}
h_{{R}}(t) = \frac{\tau_1^2 + \tau_2^2}{\tau_1 \tau_2^2}\,e^{-t/\tau_2}\,\sin(t/\tau_1)\,\Theta(t), \label{eq:hR_t}
\end{align}
with fractional contribution $f_{{R}}$, with the Heaviside step-function
$\Theta(t)$ ensuring causality.
To model the Raman effect in diamond waveguides we here use $f_{{R}}=0.20$, $\tau_1
= 4.0~\mathrm{fs}$, and $\tau_2=5.7~\mathrm{fs}$ \cite{Kardas:OE:2013}.
Using a discrete sequence of angular frequency detunings $\Omega=
\omega-\omega_0 \in 2\pi T^{-1} \mathbb{Z}$, the
expressions
\begin{subequations}\label{eq:FT}
\begin{align}
&A_\Omega(z) = \frac{1}{T} \int_{-T/2}^{T/2} A(z,t)\,e^{i\Omega t}~{\rm{d}}t,\label{eq:FT_FT}\\
&A(z,t) = \sum_{\Omega} A_\Omega(z)\,e^{-i\Omega t}, \label{eq:FT_IFT}
\end{align}
\end{subequations}
specify forward [Eq.~(\ref{eq:FT_FT})], and inverse [Eq.~(\ref{eq:FT_IFT})]
Fourier transforms, relating the field envelopes $A(z,t)$ to the spectral
envelopes $A_\Omega(z)$.
%
%
The energy of the field $A$ can be written in the form $E(z)=\hbar \sum_\Omega
n_\Omega(z) \,(\omega_0+\Omega)$, where $\hbar$ is the reduced Planck constant,
and where the dimensionless quantity $n_{\Omega}(z) \equiv T |A_{\Omega}(z)|^2/[\hbar
(\omega_0 + \Omega)]$ specifies the number of photons with energy $\hbar
(\omega_0\!+\!\Omega)$. Consequently, the total number of photons is given by
\begin{align}
C_{\rm{Ph}}(z) = \frac{2\pi}{\hbar \Delta \Omega}\sum_\Omega \frac{|A_\Omega(z)|^2}{\omega_0 + \Omega}. \label{eq:CN}
\end{align}
Let us note that the GNLS (\ref{eq:GNLS}) conserves the total number of photons
$C_{\rm{Ph}}$, but does not conserve the energy $E$ due to the Raman
interaction and self-steepening \cite{Blow:JQE:1989}. 
The numerical simulations in terms of the GNLS reported below are performed
using the variable stepsize ``conservation quantity error'' (CQE) method
\cite{Heidt:JLT:2009,Rieznik:IEEEPJ:2012,Melchert:CPC:2022,GNLStools:GitHub:2022},
with stepsize selection guided by $C_{\rm{Ph}}$.
To assess time-frequency interrelations within the field $A$
at a selected propagation distance $z$, we use the spectrogram
\cite{Melchert:SFX:2019,Cohen:IEEE:1989}
\begin{equation}
P_{S}(t,\Omega) = \frac{1}{2 \pi} \left|\int_{-T/2}^{T/2} A(z,t^\prime)h(t^\prime-t) e^{-i \Omega t^\prime}~{\rm d}t^\prime\right|^2, \label{eq:PS}
\end{equation}
wherein $h(x)=\exp(-x^2/2\sigma^2)$ is a Gaussian window function with
root-mean-square width $\sigma$, allowing to localize $A$ in time.

\begin{figure*}[t!]
\includegraphics[width=\linewidth]{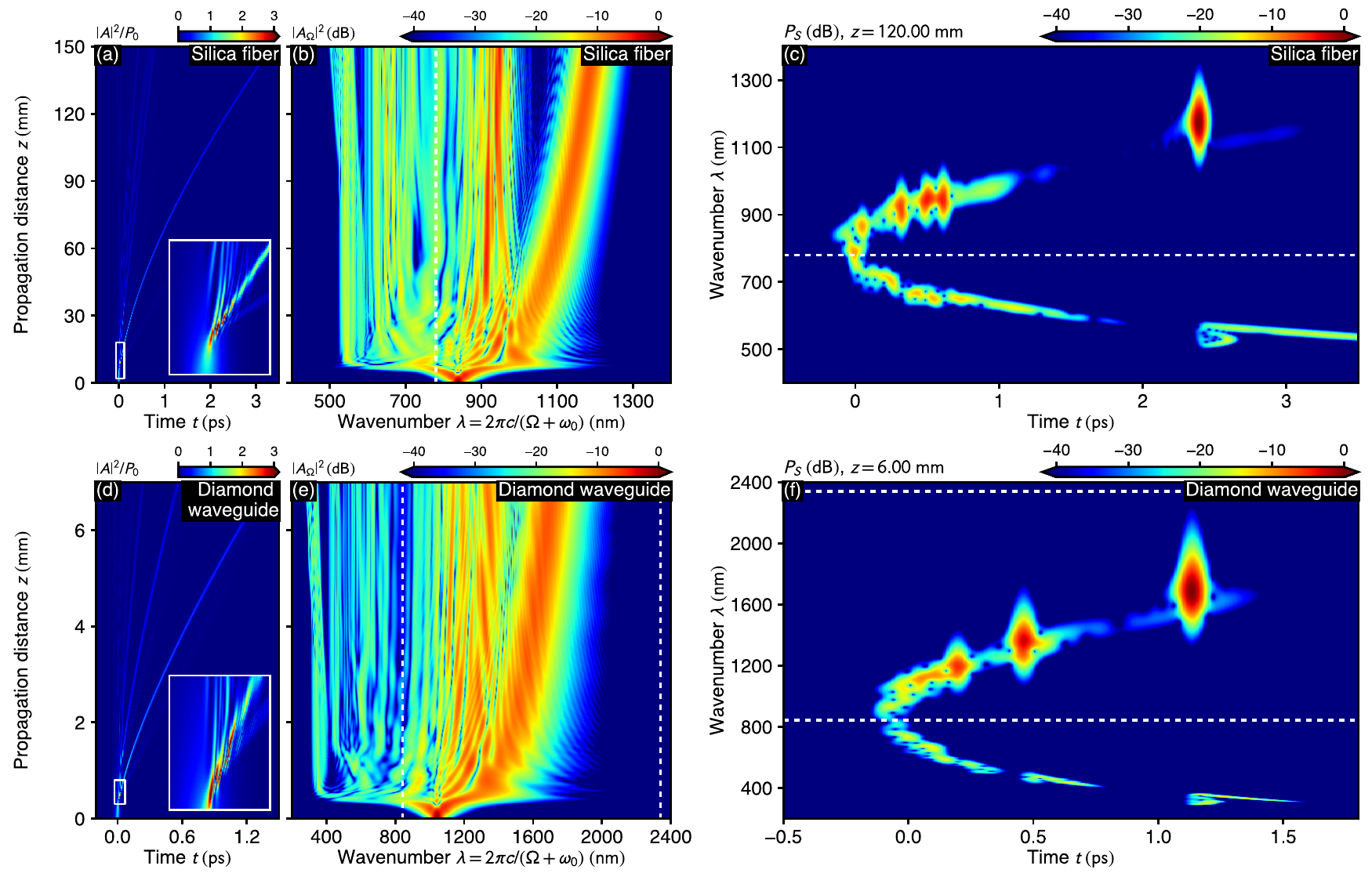}
\caption{Supercontinuum generation processes. (a-c) Results for a silica-based
optical fiber. (a) Time-domain propagation dynamics. Inset shows zoom-in on the
soliton fission process. (b) Spectral-domain propagation dynamics. Vertical
dashed line indicates zero-dispersion point at $\lambda_{\rm{Z}}\approx
780\,\mathrm{nm}$. (c) Spectrogram at $z=12\,\mathrm{cm}$ using a rms-width
$\sigma=25\,\mathrm{fs}$ to localize the field.
(d-f) Same as (a-c) for a diamond waveguide. Vertical dashed lines in (e)
indicate zero-dispersion points at $\lambda_{Z1}\approx 843\,\mathrm{nm}$ and
$\lambda_{\rm{Z2}}\approx 2340\,\mathrm{nm}$.  
\label{fig:OM:02}}
\end{figure*}

\section{Results}
\label{sec:results}

\paragraph{Supercontinuum generation}
Below we compare a supercontinuum generation process in a standard silica-based
optical fiber with with properties detailed in Ref.~\cite{Dudley:RMP:2009}, to
supercontinuum generation in a silica surrounded diamond waveguide exhibiting
the dispersion properties detailed in Fig.~\ref{fig:OM:01}.
Results of numerical simulations using initial hyperbolic-secant pulses
$A_0(t)=P_0\,{\mathrm{sech}}(t/t_0)$ with peak power $P_0$ and duration $t_0$
are shown in Fig.~\ref{fig:OM:02} (parameters are detailed below). 
%
Starting from the spectrally narrow input pulse, the interplay of linear and
nonlinear effects inherent to Eq.~(\ref{eq:GNLS}) leads to an enormous spectral
broadening. This involves soliton fission, i.e.\ the successive breakup of the
initial pulse into fundamental solitons, see the insets of
Fig.~\ref{fig:OM:02}(a) and Fig.~\ref{fig:OM:02}(d), 
for a silica-fiber and a diamond waveguide, respectively.
The pulse breakup is accompanied by the generation of dispersive waves in the
domain of normal dispersion, extending the spectrum towards the blue side.
Due to the Raman effect these solitons experience a self-frequency shift
[Figs.~\ref{fig:OM:02}(b,e)], extending the red side of the spectrum and
resulting in a deceleration of the pulses in the time domain
[Figs.~\ref{fig:OM:02}(a,d)].
Under certain conditions, the ejected solitons form strong refractive index
barriers that cannot be surpassed by quasi group-velocity matched dispersive
waves in the domain of normal dispersion, resulting in reflection processes
that further extend the blue side of the spectrum
\cite{Dudley:RMP:2009,Driben:OE:2010,Demircan:PRL:2013,Demircan:OL:2014}.
Instances of such reflection processes are visible in the spectrograms in
Figs.~\ref{fig:OM:02}(c,f).
While both supercontinuum generation processes look very similar regarding the
structure of their underlying soliton fission processes, see insets of
Figs.~\ref{fig:OM:02}(a,d), and spectrum, see Figs.~\ref{fig:OM:02}(b,e), both
occur on very different energy scales and propagation distances.
Let us note that a fundamental soliton for the silica-based optical fiber with
$\beta_2= -0.011\,\mathrm{ps^2/m}$, $\gamma=0.1\,\mathrm{W^{-1}/m}$, and, say,
$t_0=0.1\,\mathrm{ps}$, would require a peak power $P_0\approx 11\,\mathrm{W}$
and yield a soliton period $z_{S}=(\pi/2) L_D \approx 1.4\,\mathrm{m}$
(dispersion length $L_D=t_0^2/|\beta_2|$). Such a fundamental soliton would have
energy $E=2.2\,\mathrm{pJ}$.
In contrast, a diamond waveguide with $\beta_2=-0.26\,\mathrm{ps^2/m}$,
$\gamma=9.6\,\mathrm{W^{-1}/m}$, and $t_0=0.1\,\mathrm{ps}$ requires only
$P_0=2.7\,\mathrm{W}$ and exhibits $z_S\approx 0.006\,\mathrm{m}$. In this
case, $E=0.54 \,\mathrm{pJ}$, i.e.\ the energy required for the fundamental
soliton is smaller by about a factor of four.
For the supercontinuum generation process shown in Fig.~\ref{fig:OM:02}, in
case of the silica-based optical fiber, the initial pulse had peak power
$P_0=10\,\mathrm{kW}$ and duration $t_0=28.4\,\mathrm{fs}$, injected at
$\omega_0=2.260\,\mathrm{rad/fs}$ ($\lambda_0=835\,\mathrm{nm}$), corresponding
to a soliton of order $N\approx 8.7$.  
In case of the diamond waveguide, $P_0=1.66\,\mathrm{kW}$,
$t_0=20\,\mathrm{fs}$, and $\omega_0=1.82\,\mathrm{rad/fs}$ ($\lambda_0=1035\,\mathrm{nm}$), corresponding to a soliton of order $N=7$.
Let us point out that while the supercontinuum generation process in the silica
fiber develops on a lengthscale of $12\,\mathrm{cm}$, a similar dynamics in
case of the diamond waveguide unfolds on merely $6\,\mathrm{mm}$.

\begin{figure}[t!]
\includegraphics[width=\linewidth]{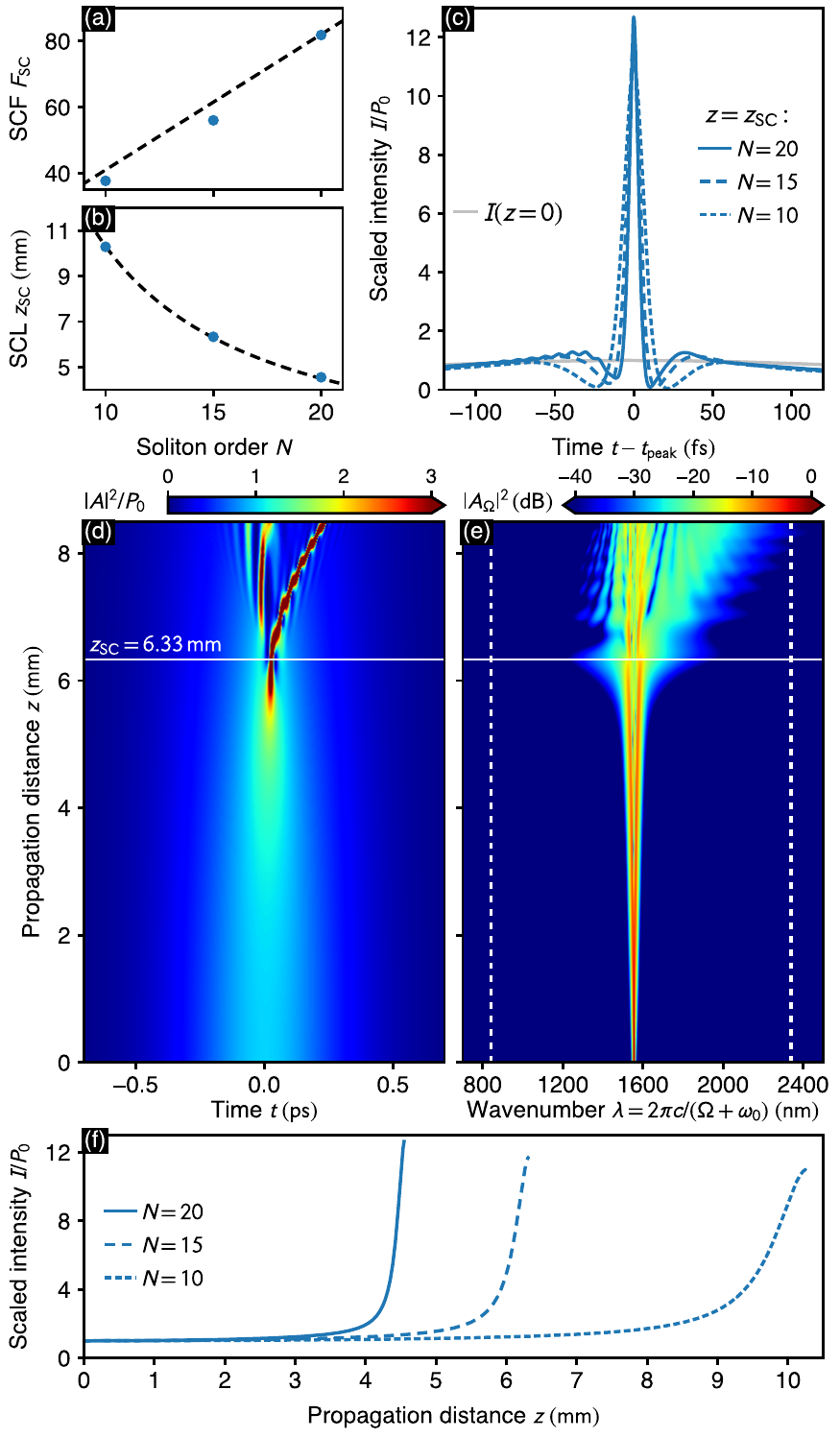}
\caption{Pulse compression scheme based on soliton dynamics.
(a) Dependence of the soliton self-compression factor (SCF) $F_{\rm{SC}}$ on
the soliton order $N$. 
(b) Dependence of the soliton self-compression length (SCL) $z_{\rm{SC}}$ on
$N$.  The scaling laws governing the dashed lines in (a,b) are detailed in the
text.  Blue dots in (a,b) are the results of numerical simulations in terms of
Eq.~(\ref{eq:GNLS}).
(c) Intensity profile at $z=z_{\rm{SC}}$ for soliton orders $N=10,\,15,\,20$, 
centered on the pulse peak position $t_{\rm{peak}}$. 
(d) Propagation dynamics in the time-domain for $N=15$, and, (e) propagation
dynamics in the spectral domain. 
Horizontal lines in (d,e) indicate the optimal self-compression distance
$z_{\rm{SC}}=6.33\,\mathrm{mm}$. Vertical dashed lines in (e) indicate
zero-dispersion points. 
(f) Variation of the pulse intensity upon propagation distance for soliton
orders $N=10,\,15,\,20$.   
\label{fig:OM:03}}
\end{figure}

\paragraph{Self-compression scheme}
The initial stage of the above supercontinuum generation processes, which is
characterized by an enormous spectral broadening, allows to identify a pulse
compression scheme, which, in the case of the diamond waveguide
[Figs.~\ref{fig:OM:02}(d-f)], proceeds on a propagation scale of less than a
millimeter.
Subsequently we discuss this initial self-compression of a higher-order
soliton, occurring in the time-domain, in more detail.
The narrowing of picosecond pulses in a silica-based single-mode optical fiber
in a domain of anomalous dispersion was demonstrated experimentally in
Ref.~\cite{Mollenauer:OL:1983}.
Therein, pulses with initial duration of $t_0\approx 4~\mathrm{ps}$
($7~\mathrm{ps}$ FWHM according to Ref.~\cite{Mollenauer:OL:1983}) were
compressed to about $1/27$ of their initial duration within a fiber of length
$320~\mathrm{m}$.
The underlying mechanism builds upon soliton effects: for a negative value of
group-velocity dispersion and for high pulse intensities, exceeding that of a
fundamental soliton, the resulting chirp across the pulse leads to pulse
narrowing upon propagation; after the pulse attains maximum compression,
soliton fission occurs [see Fig.~\ref{fig:OM:02}(d)]; the higher the initial
intensity, the smaller the propagation distance that is required to reach the
point of maximum compression.
A qualitative analysis of the propagation dynamics of solitons of high order
$N$ in terms of the common nonlinear Schrödinger equation (NLS), i.e.\
Eq.~(\ref{eq:GNLS}) with $f_R=0$ and nonzero $\beta_2<0$ only, resulted in
approximate scaling laws for the maximally attainable self-compression (SC)
factor \cite{Mollenauer:OL:1983,Oliver:OL:2021}
\begin{align}
F_{\rm{SC}} = 4.1 N, \label{eq:FSC}
\end{align}
and the optimal self-compression distance  \cite{Mollenauer:OL:1983,Oliver:OL:2021}
\begin{align}
z_{\rm{SC}} = \frac{L_D}{N} \left(0.32+\frac{1.1}{N}\right),\label{eq:zSC}
\end{align} 
where $L_D=t_0^2/|\beta_2|$ is the dispersion length of the initial pulse and
$z_{\rm{SC}}$ specifies the propagation distance at which the compression
factor $F_{\rm{SC}}$ is achieved. 

Subsequently we transfer the concept of this soliton-effect pulse compression scheme
to diamond waveguides with dispersive and nonlinear properties  detailed in
Sect.~\ref{sec:methods}.
Specifically, we consider the operating wavelength
$\lambda_0=1540\,\mathrm{nm}$ ($\omega_0 = 1.212\,\mathrm{rad/fs}$), at which
$\beta_2(\omega_0)\approx -0.59\,\mathrm{ps^2/m}$, and hyperbolic secant pulses
of duration $t_0=0.3\,\mathrm{ps}$.
Figure~\ref{fig:OM:03} summarizes the results of our numerical simulations for 
soliton orders $N=10$, $15$, and, $20$.
In Figs.~\ref{fig:OM:03}(a,b) we compare the theoretical predictions of
Eqs.~(\ref{eq:FSC}),(\ref{eq:zSC}) with simulations performed in terms of the
full GNLS, given by Eq.~(\ref{eq:GNLS}). The good agreement with the above
approximate scaling laws does not come as a surprise: 
during the initial propagation stage, i.e.\ well before soliton fission sets
in, the dynamics is well described by the common NLS.
In this regard, an important requirement is that the underlying group-velocity
dispersion exhibits a nearly flat anomalous dispersion profile, extending over
a wide wavelength range. 
Problems concerning the compression limit \cite{Demircan:PTL:2006}, due to an
overlap of the spectrally broadened pulse with the domain of normal dispersion,
are thus also reduced.
For instance, at $N=15$, we find that the initial pulse compresses down to
$5.4\,\mathrm{fs}$, resulting in a self-compression factor $F_{\rm{SC}}\approx
56$ at the optimal self-compression distance $z_{\rm{SC}}= 6.33\,\mathrm{mm}$.
Both these values are obtained from pulse propagation simulations in terms of
the GNLS~(\ref{eq:GNLS}).
Note that for the largest compression factors, the peak intensity might achieve
a $\mathrm{TW/cm^2}$ level, in which case higher-order effects such as
interband-transitions (multi-photon absorption) start to play an increasing
role. 
A visual account of the achieved compression is given in
Fig.~\ref{fig:OM:03}(c), where the pulse intensity at $z_{\rm{SC}}$ is compared
to its initial trace for the above three choices of $N$.
Similar as in Ref.~\cite{Mollenauer:OL:1983}, we observe that
for increasing $N$, an increasingly narrow central peak on top of a broad
pedestal emerges.
The propagation dynamics for the case $N=15$ is demonstrated in
Figs.~\ref{fig:OM:03}(d,e): in the propagation range up to $z_{\rm{SC}}$, the
pulse self-compression and increase in peak intensity
[Fig.~\ref{fig:OM:03}(d)], accompanied by spectral broadening
[Fig.~\ref{fig:OM:03}(e)], is clearly evident; immediately beyond
$z_{\rm{SC}}$, soliton fission sets in.
Let us emphasize that the considered diamond waveguides support the propagation
of ultrashort solitons at rather low pulse intensities. This is a special
property enabled by diamond waveguides.
Finally, let us note that upon approaching $z_{\rm{SC}}$, the peak intensity
increases at a strongly increasing rate, see Fig.~\ref{fig:OM:03}(f). 
This requires an adequate adjustment of the device length or input power to
optimally exploit this compression scheme.

\section{Discussion and conclusions}
\label{sec:discussion}

In summary, we studied the nonlinear propagation dynamics of optical
pulses in diamond waveguides in terms of the generalized nonlinear Schrödinger
equation.
Specifically, we considered a waveguide device, designed for the telecom
wavelength range with a wide domain of anomalous dispersion
\cite{Hausmann:N:2014}.
We demonstrated that the supercontinuum generation process, which for
silica-based optical fibers usually occurs on the scale of centimeters or even
meters, occurs already on the scale of millimeters.
Owing to the strong optical nonlinearity of diamond and the large negative
value of the achievable waveguide group-velocity dispersion, this process not
only occurs on much shorter propagation scales, it also requires much lower
pulse energies.
Recognizing that the propagation dynamics prior to soliton-fission
is always characterized by pulse narrowing, directly allows to transfer
a simple and efficient pulse compression scheme to the diamond platform,
promising record-breaking compression factors on chip-size propagation
distances.

This compression scheme, which is solely based on soliton effects in a
domain of anomalous dispersion, has recently been studied experimentally within
SiN waveguides \cite{Oliver:OL:2021}.
Therein, pulses with initial duration of $1.2\,\mathrm{ps}$ and soliton order
$N\approx 19$ were compressed to about $1/18$ of their initial duration within
a low-loss, dispersion engineered waveguide of $40\,\mathrm{cm}$ length.
This experimentally achieved compression factor is much below the theoretical
prediction $F_{\rm{SC}}\approx 78$ obtained via Eq.~(\ref{eq:FSC}).
Also, earlier efforts to exploit this self-compression mechanism did not yield
compression factors larger than 11
\cite{Colman:NP:2010,BlancoRedondo:NC:2014,Choi:APL:2019}.
%
%
In our case, the compression factor can simply be increased by starting with an
initial pulse duration in the picosecond range and using longer propagation
distances.
The limitation of this compression scheme is given mainly by the extend of the
domain of anomalous dispersion, supporting undisturbed soliton propagation.

While the presented study has a focus on diamond waveguides with a simple
geometry, we expect that other waveguide devices fabricated on basis of
synthetic diamond, such as angle-etched \cite{Shams:OL:2019}, and fin-shaped
structures \cite{Grote:APL:2016}, behave in a qualitatively similar manner.
The reported findings are of fundamental interest to nonlinear optics, and
provide further insight into the complex propagation dynamics of ultrashort
pulses in diamond waveguides.

\section*{Acknowledgements}
\noindent Funding: This work was supported by the Deutsche
Forschungsgemeinschaft (DFG) under Germany’s Excellence Strategy within the
Cluster of Excellence PhoenixD (Photonics, Optics, and Engineering—Innovation
Across Disciplines) [EXC 2122, Project No.\ 390833453],
and 
the European Regional Development Fund for the ‘Hannover Alliance of Research
on Diamond (HARD)’ (ZW7-85196513).



 \bibliographystyle{elsarticle-num} 
 \bibliography{references}





\end{document}